# Helicoidal magnetic structure and ferroelectric polarization in $Cu_3Nb_2O_8$

Zheng-Lu Li,[1] M.-H. Whangbo,[2] X. G. Gong,[1,*] and H. J. Xiang[1,*]

[1]*Key Laboratory of Computational Physical Sciences (Ministry of Education), State Key Laboratory of Surface Physics, and Department of Physics, Fudan University, Shanghai 200433, People's Republic of China*

[2]*Department of Chemistry, North Carolina State University, Raleigh, North Carolina 27695-8204, USA*

**ABSTRACT**

We investigate the origin of the coplanar helicoidal magnetic structure and the ferroelectric polarization in $Cu_3Nb_2O_8$ by combining first-principles calculations and our spin-induced ferroelectric polarization model. The coplanar helicoidal spin state comes from the competition between the isotropic exchange interactions, and the ferroelectric polarization from the symmetric exchange striction with slight spin canting. However, the direction of the polarization is not determined by the orientation of the spin rotation plane.

## I. INTRODUCTION

In recent years, multiferroics[1] with ferroelectric polarization ***P*** induced by a magnetic order have attracted much research interests.[1-3] In these materials the ferroelectric and magnetic orders can coexist to present a strong magnetoelectric effect, leading to their potential applications to spintronics.[4] The ferroelectric polarization ***P*** driven by spin spiral order are often explained by the spin current model[5] or equivalently the inverse Dzyaloshinskii-Moriya (DM) interaction model,[6] but these models fail in some cases,[7-10] particularly, in predicting the polarization induced by helical magnetic structure.[7,8] Recently, we developed a general model for describing ***P*** induced by a helical magnetic structure, and have successfully explained the origin of ***P*** in those multiferroics that are not accounted for by the abovementioned two models.[7,8] Johnson *et al.* have found some puzzling multiferroics where ***P*** is perpendicular to the spin rotation plane, which is not explained by the spin current model or the inverse DM interaction model.[9,10] To describe the microscopic origin of ***P***, they proposed a phenomenological ferroaxial coupling mechanism, namely, $\boldsymbol{P} = \gamma\sigma\boldsymbol{A}$, where $\gamma$ is a coupling constant, $\sigma$ represents the macroscopic chirality and ***A*** is the macroscopic axial vector that depends on the crystal structure[9,10] This mechanism indicates that ***P*** is related to both the crystal and the magnetic structures. Recently Johnson *et al.* reported that the multiferroic $CaMn_7O_{12}$ has giant polarization and used the ferroaxial mechanism to explain the origin of ***P***.[9] However, by using our general model, we showed that ***P*** originates from the symmetric exchange striction rather than spin-orbit coupling (SOC) in $CaMn_7O_{12}$.[8] Johnson *et al.* also reported another multiferroic $Cu_3Nb_2O_8$ and showed that $Cu_3Nb_2O_8$ has a non-collinear coplanar helicoidal magnetic structure with the propagation vector ***k*** = (0.4876, 0.2813, 0.2029) below a transition temperature $T \approx 24$ K, where the polarization with a magnitude of $17.8 \times 10^{-4}$ μC/cm$^2$ exists, being almost perpendicular to the spin rotation plane, with an angle ~14 ° to the vector normal to the spin rotation plane.[10] However, the polarization ***P*** was not perpendicular to ***k***, contradicting the prediction of the spin current or inverse DM interaction model, and thus is suggested to originate from the ferroaxial coupling mechanism of Johnson *et al.*

In the present work, by combining first-principles calculations with our spin-induced ferroelectric polarization model,[7,8] we extract the exchange parameters and the polarization coefficients using the mapping analysis.[7,8,11,12] The magnetic structure of $Cu_3Nb_2O_8$, including the spin rotation plane and the propagation vector ***k***, is determined by the competition between isotropic exchange interactions. Instead of SOC, the ***P*** is determined by the symmetric exchange striction with the slight spin canting due to the anisotropy of the system. The direction of ***P*** is not determined by the orientation of the spin rotation plane, in contradiction to the ferroaxial explanation.[9,10] Our study has successfully explained the origin of the magnetic structure and the ferroelectric polarization in $Cu_3Nb_2O_8$.

This work is organized as follows. In Sec. II, we describe our first-principles computational methods. In Sec. III, we present our results on the magnetic structure and the ferroelectric polarization. We demonstrate that the competition between the isotropic exchange interactions is responsible for the magnetic structure, and the symmetric exchange striction with spin canting is the origin of the ferroelectric polarization. The main conclusions are summarized in Sec. IV.

## II. CALCULATION DETAILS AND GEOMETRICAL STRUCTURES

We performed first-principles calculations based on density functional theory (DFT) by using the Vienna *ab initio* simulation package[13] (VASP) with the projector-augmented-wave method,[14,15] the generalized gradient approximation (GGA) by Perdew, Burke, and Ernzerhof[16] for exchange-correlation functional. We employed the GGA plus on-site repulsion ($U$) method[17] (GGA+$U$) to describe the strong electron correlation of Cu 3*d* orbitals, with $U = 6$ eV and $J = 1$ eV.[18,19] The plane-wave cut-off energy set 500 eV and the total energy was converged to $10^{-6}$ eV. The Berry phase method[20] was used to calculate the ferroelectric polarization. We also include the SOC effect in some calculations (GGA+$U$+SOC), to study the origin of the polarization, and in these cases, the spin direction are fixed but their magnitude are allowed to relax.

The crystal structure of $Cu_3Nb_2O_8$ has the space group of $P\bar{1}$, the lattice parameters and the atomic coordinates used for our calculations were taken from the experiment.[10] There are two non-equivalent Cu atoms, Cu1 and Cu2. In the unit cell, there is one Cu1 atom at the inversion center and the other two Cu2 atoms (marked as Cu2_1 and Cu2_2) are equivalent. The structure of $Cu_3Nb_2O_8$ is generally made up of $Cu_3O_8$ units, and in one $Cu_3O_8$ unit, the eleven atoms are almost in one plane [Fig. 1(a)]. The $Cu_3O_8$ units connect to form the staggered Cu-O chains along the *a* axis, and further the chains extend to become the Cu-O layers, separated by the non-magnetic Nb atoms in between these layers [Fig. 1(b)]. Our mapping analysis[7,8,11,12] requires the use of large supercells to avoid the interactions between spin dimers in the periodic cells, and thus we use 2×2×2, 2×2×3, 2×3×2 and 3×2×2 supercells to extract the parameters of exchange interactions and the ferroelectric polarization. The total polarization simulating the experimental magnetic structures from first-principles calculations was derived with 2×3×3 supercell. In all abovementioned supercells, we use the 2×2×2 *k*-points mesh. Convergence tests have been performed to assure the high accuracy of the energy term we are concerned with, which is the energy difference between different spin states. The spin exchange parameter, for example, $J_1$ equals -2.67 meV (see Table I) using the 2×2×2 *k*-point mesh, and -2.67 meV using the 3×3×3 *k*-point mesh. Thus, for all systems we use the 2×2×2 *k*-point mesh or even a larger k-point set to keep our results reliable.

## III. RESULTS AND DISCUSSIONS

### A. Magnetic Structure

We first consider the formation of non-collinear magnetic structure. We have considered totally ten exchange paths of the Cu pairs [Fig. 2(a)], i.e., all those Cu pair distances less than 6 Å in Cu-O chains and those less than 5 Å between the Cu-O chains and layers. The energy of exchange interactions can be written as follows:

$$E_{ex} = \sum_{<i,j>} J_{ij} \boldsymbol{S}_i \cdot \boldsymbol{S}_j \tag{1}$$

where we take $|\boldsymbol{S}_i| = 1$, and $J_{ij}$ is the effective exchange constant between spins at the *i* and *j* sites (i.e., $J_{ij}S_iS_j$). We employ the energy mapping analysis[11,12] to determine the value of the exchange interactions J values from first-principles GGA+$U$ calculations by using large enough supercells. Table I lists the Cu…Cu distances (from short to long) of the 10 spin exchange paths and their J values extracted from the mapping analysis. $J_1$ and $J_{10}$ are the interactions just in one $Cu_3O_8$ unit; $J_3$, $J_4$, $J_8$ and $J_9$ are the interactions between different $Cu_3O_8$ units in one Cu-O chain; $J_2$ and $J_7$ are the interactions between different Cu-O chains in one Cu-O layer; $J_5$ and $J_6$ are the interactions between different Cu-O layers. Considering the symmetry, $J_2$, $J_3$, $J_6$, $J_8$ and $J_{10}$ have the inversion symmetry with the inversion center locates at the mid-point of the pairs, and the other pairs have no inversion symmetry [Fig. 2(a)]. Our results show that the intrachain interactions are dominant and those interchain and interlayer ones play a minor role.

The magnetic structure of $Cu_3Nb_2O_8$ [10] shows that the three Cu atoms in one $Cu_3O_8$ unit have almost ferromagnetic (FM) spin configuration (hereafter the $Cu_3O_8$ unit in FM state is referred to as FM-u), and have almost an antiferromagnetic (AFM) spin arrangement with those in adjacent $Cu_3O_8$ units, leading to the AFM Cu-O chain (hereafter AFM-c) [Fig. 1(a)]. Let us first look at the spin structure of the Cu-O chain qualitatively. If we consider the interactions only within the $Cu_3O_8$ unit, i.e., the nearest neighbor (NN) FM $J_1$ = -2.67 meV and the next nearest neighbor (NNN) AFM $J_{10}$ = 2.78 meV, the three spins in the $Cu_3O_8$ unit would show a completely non-collinear spin state, rather than a FM arrangement found in experiment.[10] Thus, we have to take other intra-chain interactions into consideration. We consider $J_1$, $J_4$, $J_9$ and $J_{10}$ in determining the spin structure of the $Cu_3O_8$ unit and the Cu-O chain, and these four interactions are the strongest among the ten (see Table I). The arrangement of these four interactions is shown in Fig.

2(b) together with the Cu1, Cu2_1, Cu2_2 and Cu2_1' atoms. In the triangle made up of $J_1$, $J_4$ and $J_9$, the Cu1 and Cu2_1 spins would be coupled ferromagnetically not only through $J_1$, but also through $J_4$ and $J_9$. Because both $J_4$ and $J_9$ are AFM, this provide an enhanced effective FM coupling between Cu1 and Cu2_1 with $J_{FM} = J_1 - J_4 - J_9 = -5.72$ meV (see APPENDIX for details). If $|J_{FM}| > 2|J_{10}| = 5.56$ meV, the spins of the three Cu atoms in one $Cu_3O_8$ unit would have a FM arrangement to achieve the lowest energy, while the spin of Cu2_1' would be antiparallel to that of Cu1 and Cu2_1 without any constraints due to strong AFM $J_4$ and $J_9$ leading to an AFM arrangement between neighboring $Cu_3O_8$ units (see APPENDIX for details). Consequently, the magnetic structure acquires the FM-u and the AFM-c arrangements, in agreement with experiment. In fact, the smaller $J_2$ and $J_7$ would even strengthen this result in a similar manner, leading to the effective $J_{FM} = J_1 - J_4 - J_9 - J_2 - J_7 = -6.57$ meV.

To quantitatively investigate the magnetic structure of $Cu_3Nb_2O_8$, we performed Monte Carlo (MC) simulations using the J values derived from the GGA+$U$ calculations (Table I). Similar spin structures (i.e., FM-u and AFM-c) are obtained when we consider all the ten J's, when the interlayer exchanges are deleted, and when all the interlayer and interchain spin exchanges are deleted. Thus, our qualitatively analysis based only on the strongest intrachain interactions is reasonable, and the spin structure of the Cu-O chain is basically determined by the intrachain interactions. Furthermore, by considering all the ten J's, with the 4×8×8 supercell, we derive perfect the spin rotation plane from MC simulations. Given the $a$ axis parallel to the $x$ axis with the $b$ axis in the $xy$ plane, the normal vector of the spin rotation plane is specified as $\boldsymbol{n}_{expr.} = (\theta, \phi)$. Experimentally, it was found that $\theta = 75.5°$ and $\phi = 54.9°$.[10] Our MC simulations led to two normal vectors $\boldsymbol{n}_1 = (85.7°, -7.2°)$ and $\boldsymbol{n}_2 = (19.7°, -1.3°)$, which are different from each other and from the experimental values. This is expected because the J's, being isotropic interactions, cannot induce any preferred spin direction in the space.

After determining the spin rotation plane, we focus on the propagation vector $\boldsymbol{k}$. The experiment shows $\boldsymbol{k} = (k_{a*}, k_{b*}, k_{c*}) = (0.4876, 0.2813, 0.2029)$, which shows an important relation that $k_{a*} \approx k_{b*} + k_{c*} \approx 0.5$. We reproduce this relation in the following on the basis of the FM-u and AFM-c magnetic structures. We employ the notation $(n_a, n_b, n_c)$ to mark the unit cells in the direct lattice, e.g., (0, 0, 0) for the unit cell at the origin, and (1, 0, 0) for the next unit cell along the $a$ axis, etc. For the Cu1 atom, which is at the inversion center in the (0, 0, 0) cell, the corresponding $Cu_3O_8$ unit has the Cu2_1 in the (-1, 0, 0) cell and Cu2_2 in the (0, -1, -1) cell [see Fig. 2(a)]. The spins in the spiral state of $Cu_3Nb_2O_8$ are described by the expression[10]

$$\boldsymbol{S}_i = \boldsymbol{R}\cos(\boldsymbol{k}\cdot\boldsymbol{R}_L + \xi_i) + \boldsymbol{I}\sin(\boldsymbol{k}\cdot\boldsymbol{R}_L + \xi_i) \tag{2}$$

where $i$ denotes the type of the Cu atom, $\boldsymbol{R}$ and $\boldsymbol{I}$ are two orthogonal vectors determining the spin rotation plane, $\boldsymbol{R}_L = (n_a, n_b, n_c)$ represents the lattice vector in the real space, and $\xi_i$ is a relative phase ($\xi_{Cu1} = 0$, $\xi_{Cu2_1} = 1.03\pi$ and $\xi_{Cu2_2} = 1.05\pi$ from experiment).[10] Now, we introduce the in-plane spin rotation angle $\phi_i = \boldsymbol{k} \bullet \boldsymbol{R}_L + \xi_i = 2\pi n_a k_{a*} + 2\pi n_b k_{b*} + 2\pi n_c k_{c*} + \xi_i$. In this $Cu_3O_8$ unit, we have $\phi_{Cu1} = \xi_{Cu1}$, $\phi_{Cu2_1} = -2\pi k_{a*} + \xi_{Cu2_1}$, $\phi_{Cu2_2} = -2\pi k_{b*} - 2\pi k_{c*} + \xi_{Cu2_2}$. Because $\xi_i$ just represents a relative phase, we choose $\xi_{Cu1} = 0$ and thus $\phi_{Cu1} = 0$. The FM-u magnetic structure has inversion symmetry, so we have $\xi_{Cu2_1} = \xi_{Cu2_2}$, and the FM nature requires $\phi_{Cu1} = \phi_{Cu2_1} = \phi_{Cu2_2} = 0$, so that $k_{a*} = \xi_{Cu2_1}/2\pi$, and $k_{b*} + k_{c*} = \xi_{Cu2_2}/2\pi$. Because of the AFM-c spin arrangement, $k_{a*} = 0.5$, as a result, $\xi_{Cu2_1} = \xi_{Cu2_2} = \pi$, and $k_{b*} + k_{c*} = 0.5$, consistent with the experimental values. So far, the above results and analyses show that the coplanar magnetic structure (including the spin rotation plane, the FM-u and the AFM-c) is a result of the competition between the isotropic exchange interactions J's, and that the propagation vector $\boldsymbol{k}$ consistent with the experiment is derived by analyzing the crystal and magnetic structures of $Cu_3Nb_2O_8$.

### B. Ferroelectric Polarization

Now we turn to the ferroelectric polarization $\boldsymbol{P}$. To reduce the computational task, we employ the propagation vector $\boldsymbol{k}$ = (1/2, 1/3, 1/3) with a 2×3×3 supercell, and specify the spins by using Eq. (2) and the experimental $\xi_i$ [10] to simulate the experimental helicoidal magnetic structure from first-principles calculations. Our GGA+$U$ calculation gives $\boldsymbol{P}$ = (138, 294, -52)×$10^{-4}$ μC/cm$^2$ in the *xyz* coordinates, and the GGA+$U$+SOC calculation gives almost an identical result with $\boldsymbol{P}$ = (138, 302, -48)×$10^{-4}$ μC/cm$^2$. Both are substantially greater than the experimental value of 17.8×$10^{-4}$ μC/cm$^2$ in magnitude. The experiment[10] shows the angle between $\boldsymbol{P}$ and $\boldsymbol{n}_{expr.}$ (<$\boldsymbol{P}$, $\boldsymbol{n}_{expr.}$>) to be about 14°, and our calculation shows that <$\boldsymbol{P}$, $\boldsymbol{n}_{expr.}$> = 22.8° from the GGA+$U$ calculations.

The above results indicate that the polarization of $Cu_3Nb_2O_8$ basically originates from the exchange striction between Cu pairs rather than SOC because SOC does not play an important role on the polarization. This is not in support of the ferroaxial mechanism[10] because the direction of the polarization does not depend on the spin rotation plane of the magnetic structure. To see which Cu pairs are mainly responsible for the ferroelectric polarization, we apply our general spin-induced ferroelectric polarization model to this $Cu_3Nb_2O_8$ case. Our general model is illustrated in Table II. Since the $\boldsymbol{P}$ comes from the exchange striction, we have (see Table II):

$$\boldsymbol{P} = \sum_{<i,j>} \boldsymbol{P}_{ij}^{es}(\boldsymbol{S}_i \cdot \boldsymbol{S}_j)/\mathrm{V} \tag{3}$$

where $\boldsymbol{P}_{ij}^{es}$ represents the exchange striction polarization coefficients of spins at *i* and *j* sites, and V is the volume of the cell containing the pairs in the summation. We again employ the mapping analysis[7,8] to derive the exchange striction polarization coefficients from first-principles GGA+$U$ calculations, as listed in Table I. We note that those pairs with inversion symmetry have $\boldsymbol{P}^{es}$ = 0, and those without inversion symmetry have comparatively large $\boldsymbol{P}^{es}$ (Table I).

We use our model [Eq. (3)] to sum up all the pair contributions to the total polarization $\boldsymbol{P}$ with the polarization coefficients $\boldsymbol{P}^{es}$ derived from the GGA+$U$ calculations. We adopt the experimental values $\boldsymbol{k}$ = (0.4876, 0.2813, 0.2029), $\xi_{\mathrm{Cu1}}$ = 0, $\xi_{\mathrm{Cu2_1}}$ = 1.03π and $\xi_{\mathrm{Cu2_2}}$ = 1.05π to obtain the summed up result $\boldsymbol{P}$ = (30, 32, 9)×$10^{-4}$ μC/cm$^2$ (see Table III), which has the magnitude of 44.6×$10^{-4}$ μC/cm$^2$ and <$\boldsymbol{P}$, $\boldsymbol{n}_{expr.}$> = 8.6°. The obtained $\boldsymbol{P}$ is closer (see below for further discussions on the magnitude of $\boldsymbol{P}$) to the experimental value, assuring that the total $\boldsymbol{P}$ comes from the exchange striction. We now examine why the $\boldsymbol{P}$ calculated directly from first-principles calculations using $\boldsymbol{k}$ = (1/2, 1/3, 1/3) is so large. With the experimental $\xi_i$ values and $\boldsymbol{k}$ = (1/2, 1/3, 1/3), we find $\boldsymbol{P}$ = (119, 206, -22)×$10^{-4}$ μC/cm$^2$ from our model (see Table III), which is consistent with our first-principles result. This indicates that the discrepancy in the polarization between the direct first-principles results and the experimental value is due to the improper spin arrangement and the improper $\boldsymbol{k}$ value adopted in the GGA+U calculation.

It should be point out that the total polarization $\boldsymbol{P}$ is very sensitive to the propagation vector $\boldsymbol{k}$ and the relative phase $\xi_i$. We define the commensurate magnetic structure, i.e., purely FM-u and AFM-c through all crystal, with strictly $k_{a*} = k_{b*} + k_{c*} = 0.5$, $\xi_{Cu1} = 0$, and $\xi_{Cu2_1} = \xi_{Cu2_2} = \pi$. However, the real magnetic structure from experiment[10] is slightly incommensurate, with $k_{a*} = 0.4876$, $k_{b*} + k_{c*} = 0.4842$, $\xi_{Cu1} = 0$, $\xi_{Cu2_1} = 1.03\pi$ and $\xi_{Cu2_2} = 1.05\pi$. Because the crystal structure of $Cu_3Nb_2O_8$ has inversion symmetry, when $\boldsymbol{k}$ is commensurate and $\xi_{Cu2_1} = \xi_{Cu2_2} = \pi$, the magnetic structure cannot break the inversion symmetry of the crystal structure, and thus would lead no ferroelectric polarization, i.e., $\boldsymbol{P} = 0$ (see last row of Table III). Only when the $\boldsymbol{k}$ and $\xi_i$ are incommensurate (even if slightly), there would be the canting of spins, and the pair contribution to the $\boldsymbol{P}$ would not cancel out, resulting in non-zero macroscopic observable polarization. Specifically, the pairs with large $\boldsymbol{P}^{es}$, i.e., $\boldsymbol{P}_1^{es}$, $\boldsymbol{P}_4^{es}$, $\boldsymbol{P}_5^{es}$ and $\boldsymbol{P}_7^{es}$, would dominate the $\boldsymbol{P}$ due to the spin canting. However, $\boldsymbol{P}_9^{es}$ is an exception, because the contributions of the two $J_9$ pairs in one unit cell strictly cancel out due to symmetry. The incommensurate behaviors of $\boldsymbol{k}$ and $\xi_i$ may come from the perturbation of the interchain and interlayer J's, the antisymmetric DM interactions and the single-ion anisotropy.[8,11,12] These effects and hence the associated spin canting are small, but are critical in determining the magnitude of $\boldsymbol{P}$.

The above explanation needs further validation. Results from our spin-induced ferroelectric polarization model, presented in Fig. 3 (a) and (b), show how the propagation vector $\boldsymbol{k}$ and the relative phases $\xi_i$ affect the total ferroelectric polarization $\boldsymbol{P}$. The experimental values are $\boldsymbol{k} = (k_{a*}, k_{b*}, k_{c*}) = (0.4876, 0.2813, 0.2029)$, $\xi_{Cu1} = 0$, $\xi_{Cu2_1} = 1.03\pi$ and $\xi_{Cu2_2} = 1.05\pi$ [10]. We first consider the effects of $\boldsymbol{k}$. We fix two of the $k_i$ ($i = a^*, b^*, c^*$) as the experimental values, and vary the remaining component $k_i$ from 0 to 1 to obtain the three curves in Fig. 3(a). Explicitly, $\boldsymbol{P}$ could take a value in a vast range, i.e., from almost 0 to more than $1500 \times 10^{-4}$ μC/cm$^2$, owing to the largest polarization coefficients in Table I, as mentioned before. We note the three curves take the form of sine function, and this is consistent with Eq. (2). In Fig. 3(b), we keep $\boldsymbol{k}$ and $\xi_{Cu1}$ as the experimental values and see the impact to $\boldsymbol{P}$ brought about by $\xi_i$ (i = Cu2_1, Cu2_2). Again we see the curves are in the form of sine function except for the vicinity of π, where there exist fluctuations that may come from the incommensurate property of the experimental $\boldsymbol{k}$ and the other $\xi_i$. The experimentally measured polarization is $17.8 \times 10^{-4}$ μC/cm$^2$. If we take all experimental parameters for the spin structures, the predicted polarization is $44.6 \times 10^{-4}$ μC/cm$^2$ (see Table III), which is larger than the experimental value by a factor of 2.5. Several reasons may account for this discrepancy. First, the ferroelectric polarization depends sensitively on the spin structure. We note that $\xi_i$ (i = Cu2_1, Cu2_2) deviate from π very slightly, and the small deviation takes a relatively large error; as shown in Ref. 10, the $\xi_{Cu2_1}$ and $\xi_{Cu2_2}$ are 1.03(7) π and 1.05(7) π, respectively. Note that, as shown in the inset of Fig. 3(b), one can either reproduce the experimental $\boldsymbol{P}$ or even underestimate it within the error range. Second, it is very common (see for example Ref. 8) in multiferroics that the theoretical polarization is larger than the experimental value due to the small magnitude and the possible electric leakage. Third, it was found that ionic displacement may have some effects on the polarization,[21,22] which is however beyond the scope of the current work.

The above discussions are based on our polarization model. Now we show how well the spin-induced polarization model agrees with the first-principles results. As summarized in Table IV, for our first-principles calculations for the total ferroelectric polarization $\boldsymbol{P}$, we have selected several $\boldsymbol{k}$ values. Results indicate that the model give reliable results as do first principles calculations. As for the $\xi_i$ (i = Cu2_1, Cu2_2), Fig. 3(c) shows that our model generally also provides similar results as do GGA+$U$ calculations, but the results of our model are smaller than those of first-principles calculations. This may due to some longer-range interactions that we do not consider. Furthermore, our GGA+$U$ and GGA+$U$+SOC results are close, strongly indicating the $\boldsymbol{P}$ originates from the exchange striction. Taken together, our polarization model can predict similar and reliable results as do first principles calculations, and allows one to study much more configurations of $\boldsymbol{k}$ and $\xi_i$ when direct first-principles calculations cannot be afforded. All the results shown in Fig. 3 indicate that the polarization $\boldsymbol{P}$, which originates from the exchange striction, is very sensitive to both $\boldsymbol{k}$ and $\xi_i$ and can take an extremely large value, but that the experimental $\boldsymbol{P}$ is relatively rather small because the magnetic structure of $Cu_3Nb_2O_8$ is almost collinear and slightly incommensurate.

### C. Effects of Spin-Orbit Coupling

Finally, we investigate in more detail the effects of SOC on the magnetic and ferroelectric properties of $Cu_3Nb_2O_8$ by performing GGA+$U$+SOC calculations. Let us first check the isotropy of J, and we take $J_1$ and $J_4$ as examples. It is recalled that $J_1$ = -2.67 meV and $J_4$ = 2.13 meV from GGA+$U$ calculations. When SOC is considered,[12] we have $J_1^{xx}$ = -3.03 meV, $J_1^{yy}$ = -3.05 meV, $J_1^{zz}$ = -3.05 meV and $J_4^{xx}$ = 2.15 meV, $J_4^{yy}$ = 2.14 meV, $J_4^{zz}$ = 2.14 meV, showing good isotropy and consistency with our GGA+$U$ result. Given SOC as perturbation, the antisymmetric DM interaction is a second-order perturbation term and exists only when inversion symmetry absent.[23] The DM energy terms is

$$E_{DM} = \sum_{<i,j>} \boldsymbol{D}_{ij} \cdot (\boldsymbol{S}_i \times \boldsymbol{S}_j) \tag{4}$$

and $\boldsymbol{D}_{ij}$ is a vector. We again perform energy mapping analysis,[11,12] and find $\mathbf{D}_1$ = (-0.15, 0.19, 0.15) meV and $\mathbf{D}_4$ = (0.09, -0.08, 0.13) meV in the *xyz* coordinates, with the ratio |**D**/J| about 0.10 and 0.08, respectively. The comparatively small DM interaction[8,12,24] would not change the basic magnetic structure, but may determine the direction of the spin-rotation plane as a consequence of gaining more energy from it. As to the effect of SOC on $\boldsymbol{P}$, we follow the third column of Table II and calculate all $\boldsymbol{P}_{12}^{\alpha\beta}$ terms of $\vec{\vec{P}}_{12}^{int}$ for $\boldsymbol{P}_1$ (corresponding to $J_1$) as an example, using mapping analysis[7,8] with GGA+$U$+SOC calculations. We find

$$\vec{\vec{P}}_1^{int} = \begin{bmatrix} (-196,-334,150) & (-23,-40,-12) & (-16,-17,7) \\ (23,41,11) & (-197,-336,150) & (-8,-17,-3) \\ (20,20,-7) & (8,17,4) & (-195,-336,150) \end{bmatrix} \times 10^{-5}\, e\text{Å} \tag{5}$$

where the tensor corresponds to the spin represented in the *xyz* coordinates, and the diagonal terms are quite consistent with our DFT+$U$ calculated $\boldsymbol{P}_1^{es}$ = (-205, -337, 126) × $10^5$ eÅ (see Table I), and the contribution of SOC (mainly off-diagonal terms) are indeed small comparing with the effect of exchange striction. These results ensure that SOC has no obvious impact on both the magnetic structure and the ferroelectric polarization, and confirms the validity of our general spin-induced ferroelectric polarization model.[7,8]

## IV. SUMMARY

We have performed mapping analysis based on first-principles calculations to extract the exchange parameters and the polarization coefficients of $Cu_3Nb_2O_8$. The magnetic structure of $Cu_3Nb_2O_8$ originates from the competition isotropic exchange interactions and is basically described by the FM-u and AFM-c arrangements leading to a coplanar helicoidal spin spiral order. Our MC simulations lead to the magnetic structure similar to the one experimentally observed. The total polarization $\boldsymbol{P}$ of $Cu_3Nb_2O_8$ is induced by the exchange striction rather than by SOC. The magnitude of $\boldsymbol{P}$ is governed by a delicate spin canting arising from the anisotropy of the incommensurate magnetic structure, and the direction of $\boldsymbol{P}$ is not determined by the orientation of the spin rotation plane.

## ACKNOWLEDGEMENTS

This work is supported by NSF of China, the Special Funds for Major State Basic Research and the Research Program of Shanghai Municipality (Pujiang, Eastern Scholar), Foundation for the Author of National Excellent Doctoral Dissertation (FANEDD) of P. R. China.

## APPENDIX: THE MAGNETIC GROUND STATE OF THE Cu-O CHAIN

In our main text, we have qualitatively analyzed the origin of the FM-u and AFM-c magnetic structure, and now we discuss it in more detail. We have defined the effective $J_{FM}$ competing with the antiferromagnetic $J_{10}$, denoting $J_{10}$ = $J_{AFM}$ in this APPENDIX. Precisely, $J_{FM} = J_1 + J_4\mathbf{S}_{Cu1}\cdot\mathbf{S}_{Cu2_1'} + J_9\mathbf{S}_{Cu2_1}\cdot\mathbf{S}_{Cu2_1'}$, i.e., the value of $J_{FM}$ varies with the spin directions [see Fig. 2(b)]. When the magnetic structure consists of the FM-u and AFM-c arrangements, the $J_{FM}$ would become the strongest ferromagnetic coupling with $J_{FM} = J_1 - J_4 - J_9$. Because the exchange interactions are isotropic, we can fix the spin of Cu1 along the *z* axis, and that of Cu2_1 is in the *xz* plane, and thus the spin direction of the Cu2_1 is decided by a variable $\theta_1$. The spin direction of Cu2_2 is decided by $\theta_2$ and $\varphi_2$. For any $\theta_1$ and $\theta_2$, to lower the energy of spin exchange between Cu2_1 and Cu2_2, we have $\varphi_2 = 0$ or $\pi$, i.e., all three spins are in the *xz* plane.

Considering the periodicity, the energy of the Cu-O chain can be written as

$$\begin{aligned}\mathrm{E} &= J_{FM}\boldsymbol{S}_{Cu1}\cdot\boldsymbol{S}_{Cu2_1} + J_{FM}\boldsymbol{S}_{Cu1}\cdot\boldsymbol{S}_{Cu2_2} + J_{AFM}\boldsymbol{S}_{Cu2_1}\cdot\boldsymbol{S}_{Cu2_2}\\ &= J_{FM}cos\theta_1 + J_{FM}cos\theta_2 + J_{AFM}cos\theta_{12}\end{aligned} \tag{A1}$$

where $\theta_{12} = \min(\theta_1 + \theta_2, 2\pi - \theta_1 - \theta_2)$ [min(A, B) equals to the smaller one between the two numbers A and B]. We rewrite Eq. (A1)

$$\begin{aligned}\mathrm{E} &= J_{FM}cos\theta_1 + J_{FM}cos\theta_2 + J_{AFM}cos(\theta_1 + \theta_2)\\ &= J_{FM}cos\theta_1 + J_{FM}cos\theta_2 + J_{AFM}(cos\theta_1 cos\theta_2 - sin\theta_1 sin\theta_2)\end{aligned} \tag{A2}$$

We make $a = cos\theta_1$, $b = cos\theta_2$, and thus

$$\mathrm{E} = J_{FM}(a+b) + J_{AFM}[ab - \sqrt{(1-a^2)(1-b^2)}] \tag{A3}$$

Note that $J_{AFM} > 0$ and the basic inequality that $xy \le \frac{x^2+y^2}{2}$, where $x$ and $y$ are any real numbers. Thus, we have

$$E \geq J_{FM}(a+b) + J_{AFM}\left[ab - \frac{2-a^2-b^2}{2}\right] = J_{FM}(a+b) + J_{AFM}\left[\frac{(a+b)^2}{2} - 1\right]$$

(A4)

where the equality is valid when $a = b$. If we take $t = (a + b)$ as a variable, the right hand side of Eq. (A4) is a quadratic function $\frac{J_{AFM}}{2}t^2 + J_{FM}t - J_{AFM}$. Note that $-2 \leq t \leq 2$, $\frac{J_{AFM}}{2} > 0$, $-\frac{J_{FM}}{J_{AFM}} > 0$, and the minimum occurs at $t = \min(-\frac{J_{FM}}{J_{AFM}}, 2)$. Thus, when $-\frac{J_{FM}}{2J_{AFM}} < 1$, the three spins form a non-collinear spiral state. When $-\frac{J_{FM}}{2J_{AFM}} \geq 1$, $t = 2$ and $cos\theta_1 = cos\theta_2 = 1$ (i.e., $\theta_1 = \theta_2 = 0$), the lowest energy results with the three spins in FM arrangement. When $J_{FM} = J_1 - J_4 - J_9$ (i.e., the strongest ferromagnetic coupling), we have $|J_{FM}| = |J_1 - J_4 - J_9| = 5.72$ meV > $2|J_{AFM}| = 5.56$, and at this time, we have the AFM-c arrangement. Note that any other spin configuration cannot lead to a lower energy, which confirms that the FM-u and AFM-c arrangements form the magnetic ground state of the Cu-O chain.

*xggong@fudan.edu.cn
*hxiang@fudan.edu.cn

**Table I.** The exchange interactions and the exchange striction polarization coefficients derived from first-principles calculations with the mapping methods.

| Path | Pair | Type[a] | Pair distance ( Å) | J (meV) | $\mathbf{P}^{es}$ [$10^{-5}$ (e ·Å)] |
|---|---|---|---|---|---|
| 1 | Cu1-Cu2 | SE | 2.918 | -2.67 | (-205, -337, 126) |
| 2 | Cu2-Cu2 | SE-I | 3.071 | 0.59 | (0, 0, 0) |
| 3 | Cu2-Cu2 | SE-I | 3.110 | -0.10 | (0, 0, 0) |
| 4 | Cu1-Cu2 | SE | 3.123 | 2.13 | (-174, -243, -96) |
| 5 | Cu1-Cu2 | SSE | 4.474 | -0.12 | (9, 4, -10) |
| 6 | Cu2-Cu2 | SSE-I | 4.849 | -0.03 | (0, 0, 0) |
| 7 | Cu1-Cu2 | SSE | 4.914 | 0.26 | (28, 22, 23) |
| 8 | Cu1-Cu1 | SSE-I | 5.183 | 0.00 | (1, 1, 0) |
| 9 | Cu2-Cu2 | SSE | 5.183 | 0.92 | (-69, -92, -26) |
| 10 | Cu2-Cu2 | SSE-I | 5.837 | 2.78 | (0, 0, 0) |

[a] SE represents superexchange, SSE super-superexchange. The exchange pair with inversion symmetry is indicated by adding –I.

**Table II.** Explaining the ferroelectric polarization induced by a helical magnetic structure using our spin-induced ferroelectric polarization model. In general, the polarization $\boldsymbol{P}$ has two origins, i.e., the exchange striction and SOC. Under different situations, our model is reduced to simpler forms.

<table>
<tr><td colspan="3">$$\boldsymbol{P}_{12}(\boldsymbol{S}_1,\boldsymbol{S}_2) = \sum_{\alpha\beta} \boldsymbol{P}_{12}^{\alpha\beta} S_{1\alpha} S_{2\beta} = (S_{1x}, S_{1y}, S_{1z}) \begin{pmatrix} \boldsymbol{P}_{12}^{xx} & \boldsymbol{P}_{12}^{xy} & \boldsymbol{P}_{12}^{xz} \\ \boldsymbol{P}_{12}^{yx} & \boldsymbol{P}_{12}^{yy} & \boldsymbol{P}_{12}^{yz} \\ \boldsymbol{P}_{12}^{zx} & \boldsymbol{P}_{12}^{zy} & \boldsymbol{P}_{12}^{zz} \end{pmatrix} \begin{pmatrix} S_{2x} \\ S_{2y} \\ S_{2z} \end{pmatrix} = \boldsymbol{S}_1 \vec{\vec{P}}_{12}^{int} \boldsymbol{S}_2$$</td></tr>
<tr><td>Exchange striction[8]</td><td colspan="2">Spin-orbit coupling (SOC)</td></tr>
<tr><td rowspan="2">Any rotation is allowed, so the diagonal terms of $\vec{\vec{P}}_{12}^{int}$ are equal and the off-diagonal ones are zero:<br>$\boldsymbol{P}_{12} = \boldsymbol{P}_{12}^{es}(\boldsymbol{S}_1 \cdot \boldsymbol{S}_2)$<br>$\boldsymbol{P}_{12}^{es} = 0$ if the spin dimer has inversion symmetry.</td><td>With inversion symmetry[7]</td><td>With no inversion symmetry</td></tr>
<tr><td>$\vec{\vec{P}}_{12}^{int}$ reduces to an antisymmetric tensor.<br>Generalized spin current model:<br>$\boldsymbol{P}_{12} = \boldsymbol{M}(\boldsymbol{S}_1 \times \boldsymbol{S}_2)$<br>where $\boldsymbol{M}$ is a matrix.</td><td rowspan="2">All $\boldsymbol{P}_{12}^{\alpha\beta}$ terms of $\vec{\vec{P}}_{12}^{int}$ are needed from first-principles.</td></tr>
<tr><td>$CaMn_7O_{12}$, $Cu_3Nb_2O_8$</td><td>$MnI_2$</td></tr>
</table>

**Table III.** The total polarization derived from our spin-induced ferroelectric polarization model for the case of the exchange striction, with different propagation vector $\boldsymbol{k}$ and relative phases. $\xi_{\mathrm{Cu1}}$ = 0 is fixed in this table.

| Spin orientation $\boldsymbol{S}_i = \boldsymbol{R}\cos(\boldsymbol{k}\cdot\boldsymbol{R}_L + \xi_i) + \boldsymbol{I}\sin(\boldsymbol{k}\cdot\boldsymbol{R}_L + \xi_i)$ | | | Polarization $\boldsymbol{P} = \sum_{<i,j>} \boldsymbol{P}_{ij}^{es}(\boldsymbol{S}_i\cdot\boldsymbol{S}_j)/\mathrm{V}$ |
|---|---|---|---|
| $\boldsymbol{k} = (k_{a*}, k_{b*}, k_{c*})$ | $\xi_{\mathrm{Cu2_1}}$ / π | $\xi_{\mathrm{Cu2_2}}$ / π | $\boldsymbol{P}$ ($\times 10^{-4}$ μC/cm$^2$) |
| (0.4876, 0.2813, 0.2029) | 1.03 | 1.05 | (30, 32, 9) |
| (0.4876, 0.2813, 0.2029) | 1.0 | 1.0 | (10, 10, 4) |
| (1/2, 1/3, 1/3) | 1.03 | 1.05 | (119, 206, -22) |
| (1/2, 1/3, 1/3) | 1.0 | 1.0 | (165, 283, -30) |
| (1/2, 1/4, 1/4) | 1.03 | 1.05 | (13, 12, 3) |
| (1/2, 1/4, 1/4) | 1.0 | 1.0 | (0, 0, 0) |

**Table IV.** Values of $\boldsymbol{P}$ obtained from the first principles and model calculations for various $\boldsymbol{k}$ with $\xi_{Cu1}$ = 0, $\xi_{Cu2_1}$ = 1.03$\pi$ and $\xi_{Cu2_2}$ = 1.05$\pi$. $\boldsymbol{P}$ varies $\boldsymbol{k}$ drastically, showing that the model can predict results consistent with first-principles calculations.

| Propagation vector $\boldsymbol{k}$ | Polarization $\boldsymbol{P}$ ($\times10^{-4}$ μC/cm$^2$) | | |
|---|---|---|---|
| | GGA+$U$ | GGA+$U$+SOC | Model |
| (1, 1/2, 1/2) | (5, 8, -5) | (5, 7, -3) | (-1, -1, 2) |
| (1/2, 1, 1/2) | (2, 6, -2) | (3, 4, -4) | (3, 5, -1) |
| (1/2, 1/2, 1) | (3, 5, -1) | (5, 6, -4) | (3, 5, 0) |
| (1/2, 1/2, 1/2) | (790, 1141, -125) | (806, 1151, -120) | (732, 1179, -91) |
| (1/2, 1/2, 1/3) | (567, 898, -146) | (563, 904, -106) | (498, 804, -63) |
| (1/2, 1/3, 1/2) | (523, 953, -173) | (532, 989, -134) | (499, 805, -62) |
| (1/3, 1/2, 1/2) | (538, 727, -39) | (535, 754, -22) | (511, 823, -69) |
| (1/2, 1/3, 1/3) | (138, 294, -52) | (138, 302, -48) | (119, 206, -22) |

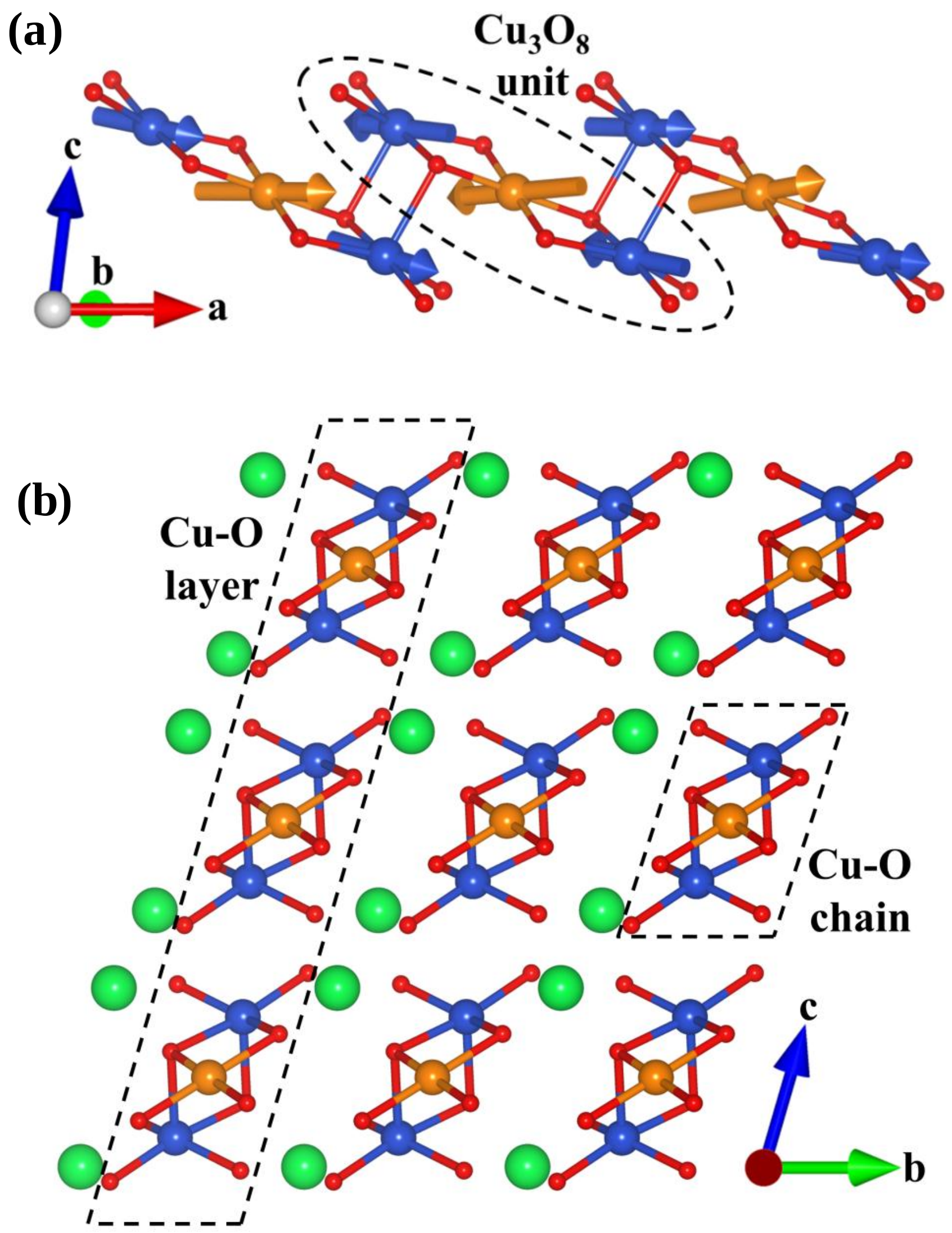

**FIG. 1** (Color online). Crystal structure of $Cu_3Nb_2O_8$. (a) The $Cu_3O_8$ units connect to form the Cu-O chain. The Cu1 atoms that are at the inversion center are in brown, and the Cu2 atoms are in blue. The magnetic structure is approximately given by the FM-u and AFM-c arrangements. (b) Cu-O chains stacked to form Cu-O layers, separated by non-magnetic Nb atoms. The Nb-O bonds are not shown.

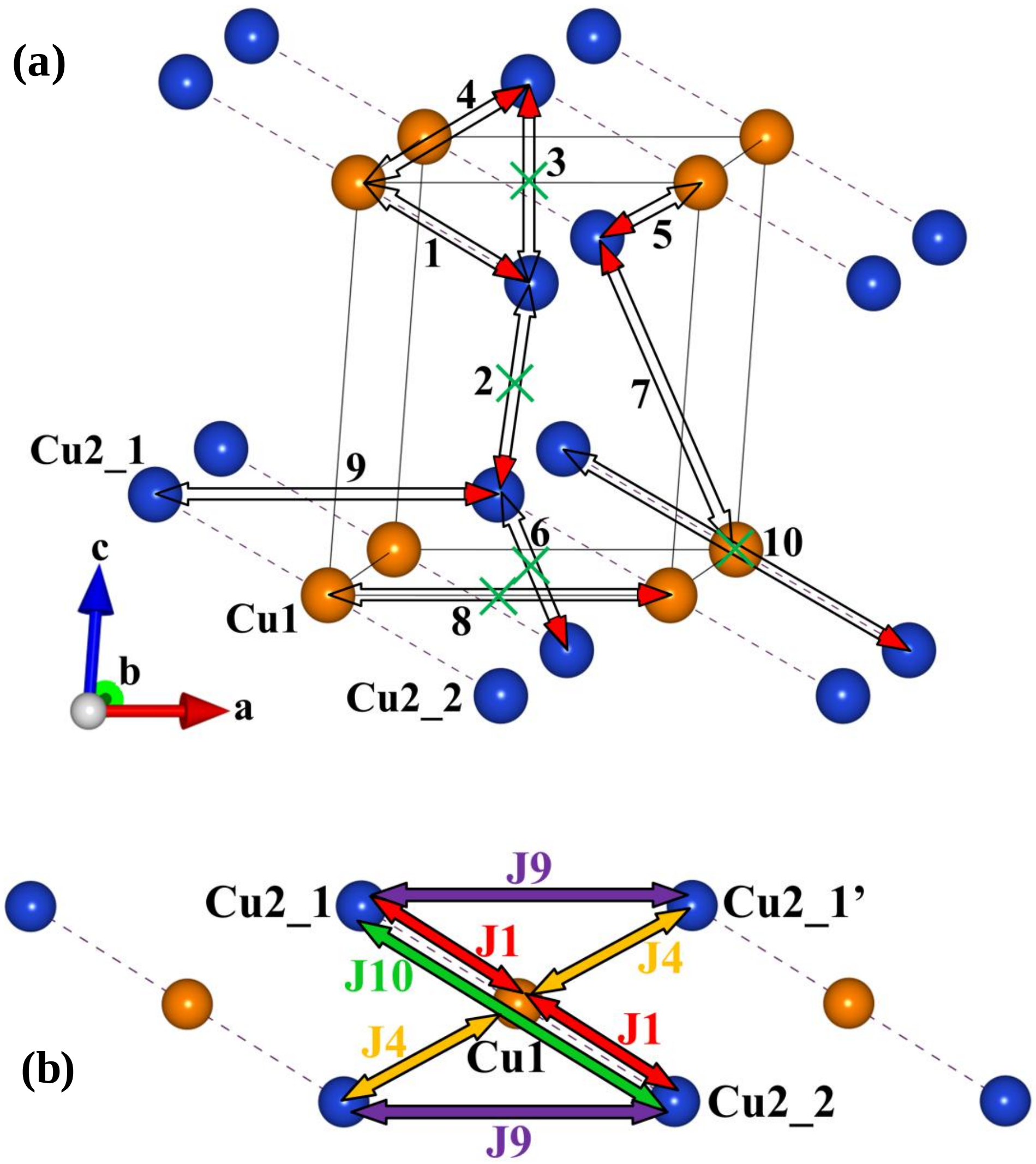

**FIG. 2** (Color online). (a) The ten spin exchange paths. The dashed line represents one $Cu_3O_8$ unit. The empty arrow points to the *i*th atom and the red solid arrow the *j*th atom. The inversion centers of those pairs with inversion symmetry are marked as green cross. (b) The main intrachain interactions leading to the FM-u and AFM-c magnetic structure, which is determined not only by the intra-trimer FM $J_1$ and AFM $J_{10}$ but also by the inter-trimer AFM $J_4$ and $J_9$.

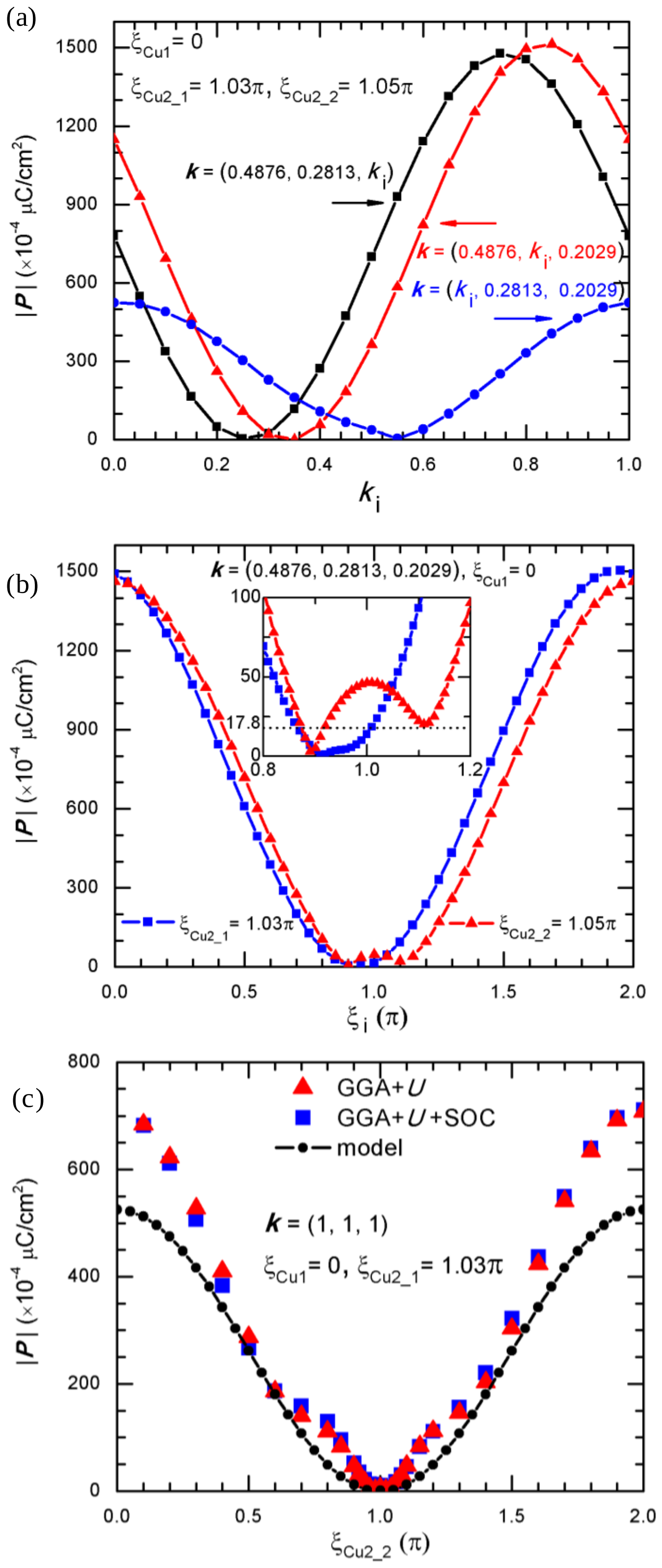
(a)
ξCu1= 0
ξCu2_1= 1.03π, ξCu2_2= 1.05π
k = (0.4876, 0.2813, ki)
k = (0.4876, ki, 0.2029)
k = (ki, 0.2813, 0.2029)
|P| (×10^-4 μC/cm^2)
ki
(b)
k = (0.4876, 0.2813, 0.2029), ξCu1= 0
17.8
ξCu2_1= 1.03π
ξCu2_2= 1.05π
|P| (×10^-4 μC/cm^2)
ξi (π)
(c)
GGA+U
GGA+U+SOC
model
k = (1, 1, 1)
ξCu1= 0, ξCu2_1= 1.03π
|P| (×10^-4 μC/cm^2)
ξCu2_2 (π)

**FIG. 3** (Color online). (a) The effect of the propagation vector ***k*** on the magnitude of ***P*** predicted from our model using the ferroelectric polarization coefficients derived from the first-principles calculations. We take the experimental $\xi_i$ and fix two components of ***k***, to see how the third component affects ***P***. Results indicate that ***P*** varies in a wide range as a function of ***k***. The experimental magnetic structure is almost collinear so that ***k*** satisfies $k_{a*} \approx k_{b*} + k_{c*} \approx 0.5$, leading to a small ***P***. (b) The effect of the relative phase $\xi_i$ to ***P***, indicated by the polarization model. Results show that ***P*** is very sensitive to a small change in $\xi_i$, given the experimental value of $17.8\times10^{-4}$ μC/cm$^2$ (see inset). (c) The effect of $\xi_{Cu2_2}$ on ***P*** obtained from the first-principles and the model calculations for a simple system with ***k*** = (1, 1, 1), for which both the GGA+*U* and the GGA+*U*+SOC calculations can be performed without much computation task. The DFT and the model calculations produce similar results, indicating the correctness of the model. See Table IV for the effect of ***k*** on ***P***.